\documentclass[useAMS,usenatbib]{mn2e}
\usepackage{amssymb,amsmath}
\usepackage{graphicx}
\usepackage{natbib}
\usepackage{journal_shortcuts}
\newcommand{\Sun}{_{\sun}}

\newcommand{\Gal}{_{\mathrm{gal}}}
\newcommand{\ICM}{_{\mathrm{ICM}}}
\newcommand{\Gas}{_{\mathrm{gas}}}

\newcommand{\Ram}{_{\mathrm{ram}}}
\newcommand{\KH}{_{\mathrm{KH}}}
\newcommand{\Grav}{_{\mathrm{grav}}}
\newcommand{\degree}{^o}
\newcommand{\K}{\,\textrm{K}}
\newcommand{\Kpc}{\,\textrm{kpc}}
\newcommand{\Mpc}{\,\textrm{Mpc}}

\newcommand{\Yr}{\,\textrm{yr}}

\newcommand{\Myr}{\,\textrm{Myr}}

\newcommand{\Kms}{\,\textrm{km}\,\textrm{s}^{-1}}
\newcommand{\Cm}{\,\textrm{cm}}
\newcommand{\Erg}{\,\textrm{erg}}
\newcommand{\ccm}{\,\textrm{cm}^{-3}}
\newcommand{\gccm}{\,\textrm{g}\,\textrm{cm}^{-3}}
\newcommand{\Presunit}{\,\textrm{erg}\,\textrm{cm}^{-3}}

\title[Ram pressure stripping of disc galaxies orbiting in clusters]%
{Ram pressure stripping of disc galaxies orbiting in clusters. I. Mass and
  radius of the remaining gas disc}

\author[E. Roediger and M. Br\"uggen]%
{Elke Roediger%
\thanks{E-mail:
e.roediger@jacobs-university.de, m.brueggen@jacobs-university.de}
and  
Marcus Br\"uggen\footnotemark[1]%
\\
Jacobs University Bremen, P.O. Box 750\,561, 28725 Bremen,
Germany}

\begin{document}

\date{Accepted. Received; in original form }

\pagerange{\pageref{firstpage}--\pageref{lastpage}} \pubyear{2007}

\maketitle

\label{firstpage}

\begin{abstract}
  We present the first 3D hydrodynamical simulations of ram pressure stripping
  of a disc galaxy orbiting in a galaxy cluster. Along the orbit, the ram
  pressure that this galaxy experiences varies with time. In this paper, we
  focus on the evolution of the radius and mass of the remaining gas disc and
  compare it with the classical analytical estimate proposed by
  \citet{gunn72}. We find that this simple estimate works well in predicting
  the evolution of the radius of the remaining gas disc. Only if the ram
  pressure increases faster than the stripping timescale, the disc radius
  remains larger than predicted.  However, orbits with such short
  ram pressure peaks are unlikely to occur in other than compact clusters. 
   Unlike the radius evolution, the mass loss history for
  the galaxy is not accurately described by the analytical estimate.
  Generally, in the simulations the galaxy loses its gas more slowly than
  predicted.
\end{abstract}


%
%
%
%
%
\section{Introduction}
%
Galaxies populate different environments, ranging from isolated field regions
to dense galaxy clusters. Depending on the environment, the properties of
galaxies, especially disc galaxies,
change: In denser regions, disc galaxies tend to contain less neutral gas, show
a weaker star formation activity and redder colours than galaxies in sparse
regions (e.g.~\citealt{vangorkom04,goto03b} and references therein).  

Ram pressure stripping (RPS), i.e. the removal of a galaxy's gas disc due to
its motion through the intracluster medium (ICM), is one of the likely
candidates to explain these features. \citet{gunn72} (GG72) proposed that for
galaxies moving face-on through the ICM the success or failure of RPS can be
predicted by comparing the ram pressure with the galactic gravitational
restoring force per unit area.  Hydrodynamical simulations of RPS presented so
far (\citealt{abadi99}, \citealt{quilis00}, \citealt{schulz01} (SS01),
\citealt{marcolini03} (MBD03), \citealt{roediger05} (RH05),
\citealt{roediger06} (RB06)) suggest that this analytical estimate does indeed
do a fair job as long as the galaxies are not moving close to edge-on.
However, in all these simulations the model galaxy was not exposed to a
varying ICM wind, as one expects for cluster galaxies, but to a constant one.
A few 2D simulations of spherical galaxies on radial orbits in clusters
(\citealt{lea76}, \citealt{takeda84}, \citealt{acreman03}) exist, where the
early ones were run with low resolution. \citet{toniazzo01} presented 3D
hydro-simulations for a spherical galaxy on a cluster orbit. Vollmer et al.
(e.g.~1999, 2000, 2001, 2001a,
2003\nocite{vollmer01a,vollmer00,vollmer01,vollmer99,vollmer03}) have
simulated the cluster passage of disc galaxies with a sticky-particle code,
where the ram pressure is modelled as an additional force on the disc gas
particles.  Their work shows the importance of the modelling of realistic
orbits, e.g. to determine the stripping history of individual galaxies.
However, the sticky-particle code cannot model hydrodynamical effects such as
instabilities that have been shown to play a role by the hydro-simulations.
Additionally, the work of SS01, RH05 and
RB06\nocite{schulz01,roediger05,roediger06} has shown that the gas removal
from the galactic potential does not occur instantaneously, but that it takes
some time until the gas is accelerated enough to become unbound from the
galaxy's potential. Thus, in cases of short ram pressure peaks, this time
delay may play an important role.

Here we present the first 3D hydrodynamical simulations of RPS a of disc
galaxy orbiting in galaxy cluster. We focus on the question, in how far a
time-dependent version of the analytical estimate based on the GG72
criterion can describe the evolution of the remaining gas disc.

%

\section{Method}
%
We model the flight of a disc galaxy through a galaxy cluster. The galaxy
starts at a position $\sim 1$ to $2\Mpc$ (depending on orbit) from the cluster
centre with a given initial velocity. We use analytical
potentials for the galaxy and the cluster, as this reduces computational costs
significantly. Given the high velocities of cluster galaxies, the tidal effect
on the galaxy is expected to be small.  The work of Moore et al. (1996, 1998,
1999)\nocite{moore96,moore98,moore99} demonstrated that only harassment, i.e.
the cumulative effect of frequent close high velocity encounters between
cluster galaxies and the overall tidal field of the cluster affects cluster
galaxies seriously. Thus, we expect that our treatment yields reasonable
results. 

The orbit of the galaxy is determined by integrating the motion of a point
mass through the gravitational potential of the cluster.  In the course of the
simulation, the position of the galaxy potential is shifted along this orbit.

\subsection{Code} \label{sec:code}
\begin{figure*}
 \centering
\includegraphics[width=\textwidth]{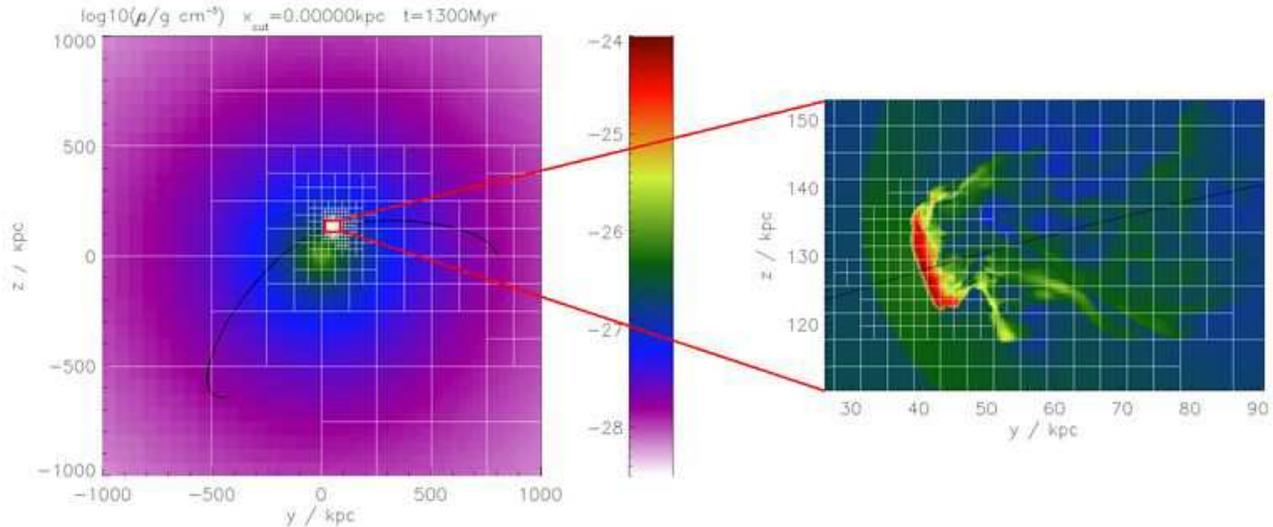}
 \caption{Colour-coded gas density in the
   orbital plane for run C2-SM-SLW-MFM (here with best resolution
   $0.25\Kpc$). The lhs picture shows a slice through the complete
   simulation box, the rhs a blow-up of the region around the galaxy. The
   black line marks the galaxy's orbit. The white rectangles show the adaptive
   grid: Each rectangle marks the boundaries of one block, each block consists
   of $8^3$ grid cells.}
 \label{fig:slice_dens}
 \end{figure*}

The simulations were performed with the FLASH code (\citealt{fryxell00}), a
multidimensional adaptive mesh refinement hydrodynamics code.  It solves the Riemann problem
on a Cartesian grid using the Piecewise-Parabolic Method (PPM).  The
simulations presented here are performed in 3D. The gas obeys a polytropic
equation of state with an adiabatic index of $\gamma=5/3$.  The size of the simulation
box is chosen such that the galaxy's orbit during the simulation time (3 Gyr)
fits into the grid. Depending on the orbit, the size of the simulation box
ranges between $(2\Mpc)^3$ and $2\times 5\times 2 \Mpc^3$. For an example see Fig.~\ref{fig:slice_dens}.  All boundaries are reflecting.
 
 The coarsest refinement level has a resolution of $\Delta x =62.5\Kpc$. For most runs,
 we use 8 levels of refinement, i.e. the best resolution is $\Delta x =0.5\Kpc$. In
 addition to the standard density and pressure gradient criteria, our
 user-defined refinement criteria enforce maximal refinement on the galactic
 gas disc and enforce stepwise de-refinement with increasing distance to the
 galaxy in the following fashion:
\begin{itemize}
\item We always refine a sphere of radius $50\Kpc$ around the galaxy centre to at
  least $\Delta x=3.9\Kpc$ (5 refinement levels).
\item We always use the maximum refinement level for the galactic disc region,
  which is defined as the cylinder with radius $27\Kpc$ and thickness $6\Kpc$
  around galactic centre, where the density larger than $10^{-27}\gccm$.
\item We limit the refinement outside $50\Kpc$ around galaxy centre to $\Delta x
  =2\Kpc$ (6 refinement levels).
\item We limit the refinement outside $150\Kpc$ around galaxy centre to $\Delta
  x=15.6\Kpc$ (3 refinement levels).
\item Additionally, we always refine a sphere of radius $100\Kpc$ around cluster centre to $\Delta
  x=7.8\Kpc$ (4 refinement levels).
\end{itemize}
Our user-defined de-refinement rules overwrite the standard pressure and
density gradient criteria, but our user-defined refinement criteria overwrite
the de-refinement criteria. The FLASH code also ensures that neighbouring
blocks differ by maximal one refinement level.

A detailed discussion of the influence of the resolution on our results is given in Sect.~\ref{sec:resolution}.

\subsection{Model galaxy}
The galaxy model is the same as in RB06\nocite{roediger06}, i.e. a massive
spiral with a flat rotation curve at $200\Kms$. It consists of a dark matter
halo, a stellar bulge, a stellar disc and a gaseous disc. All non-gaseous
components just provide the galaxy's potential and are not evolved during the
simulation. For a description of the individual components and a list of
parameters please refer to RB06\nocite{roediger06}.

\begin{figure}
 \centering\resizebox{\hsize}{!}{\includegraphics[angle=0]{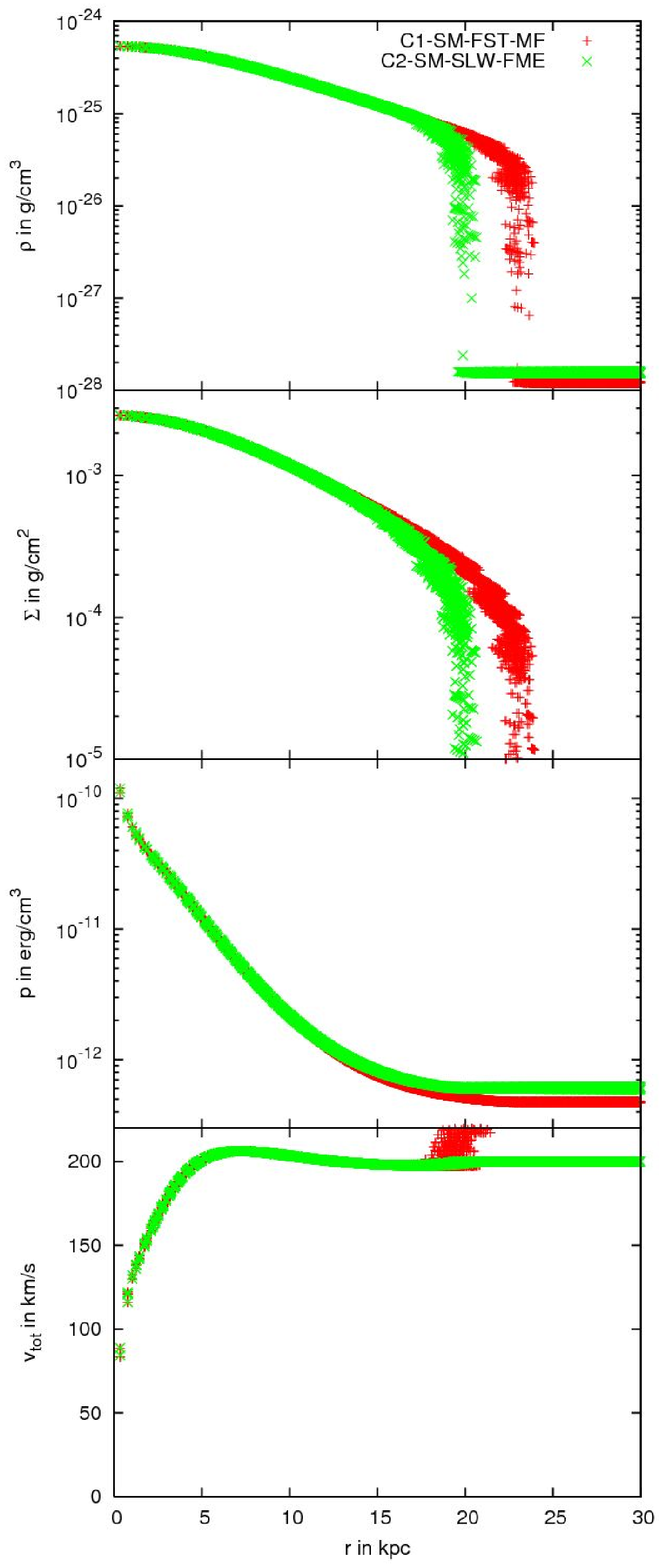}}
 \caption{Radial profiles of the density, $\rho$, pressure, $p$,
   and total velocity, $v_{\mathrm{tot}}$, in the galactic plane for the
   initial model for two different runs (see legend). Also the radial profile
   for the projected ISM surface density, $\Sigma$, is shown. One dot for
   each cell in the galactic plane is plotted. In the bottom panel, the total
   velocity in the galaxy's rest frame is shown. Velocities inside the disc
   radius (as set by e.g. the density profile) are due to the galaxy's
   rotation, while velocities outside the disk radius are due to the galaxy's
   motion through the ICM.}
 \label{fig:initial_profiles}
 \end{figure}

Initially, the galaxy is set in pressure equilibrium with the surrounding
ICM. For the case of a homogeneous ICM density and pressure, this task is
straightforward. In our case, however, there is the difficulty that both,
pressure and density of surrounding ICM, vary with distance from the cluster centre.
Consequently, the ICM pressure at the "surface'' of the galaxy is not the same
everywhere. Therefore, we use the following procedure for setting the galaxy in pressure
equilibrium with the surrounding ICM:
\begin{enumerate}
\item We chose values for the ICM density and pressure somewhat lower than the
  values at the galaxy's initial position. 
\item We calculate the density, pressure and rotation velocity in the galaxy's
  gas disc such that the gas disc is in pressure equilibrium with a
  homogeneous ambient gas of density and pressure chosen in (i).
\item We now need to ensure a continuous transition from the galactic disc to
  the cluster ICM. Therefore, we compare the pressure calculated in (ii) with
  the ICM pressure in each grid cell. Inside the gas disc the pressure
  calculated in (ii) is larger than the ICM pressure. However, at the
  transition to the ambient gas used in (ii), the pressure set in (ii) will be
  lower than the ICM pressure. In such grid cells, we substitute the values
  from (ii) with ICM values. This ensures that the pressure distribution is
  continuous everywhere. However, the density distribution shows a jump between the ICM
  and the ISM. Rotation velocity plus initial orbital velocity is only
  set for ISM cells.
\end{enumerate}
Figure~\ref{fig:initial_profiles} displays radial profiles for the density,
projected ISM surface density, pressure, and total velocity in the galactic
plane for two runs. 
The galaxy's initial position in these two runs differs
(see Table~\ref{tab:runs}). Therefore, the gas discs have to be in pressure
equilibrium with different external pressures and thus differ slightly in
radius. Moreover, the radius of each gas disc varies slightly with azimuthal angle, because the pressure in the ICM surrounding the galaxy is
not homogeneous. However, at least in the inner $10\Kpc$, the profiles of
different runs are indistinguishable.

\subsection{Cluster model}
\begin{figure}
 \centering\resizebox{0.8\hsize}{!}{\includegraphics[angle=-90]{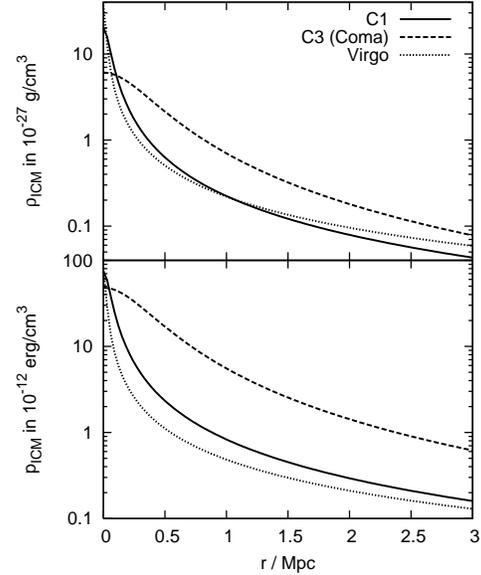}}
 \caption{Density and pressure profiles for model clusters C1 and C3. The
   profiles for cluster C2 are a factor of 2 lower than the ones of cluster
   C1. For comparison,
   the profiles for the Virgo cluster are shown.}
 \label{fig:cluster_profiles}
 \end{figure}

%
The cluster is set in hydrostatic equilibrium, where the density of the ICM follows a $\beta$-profile,
%
\begin{equation}
\rho\ICM (r) = \rho\ICM{}_0 \left[ 1+\left( \frac{r}{R\ICM}  \right)^2  \right]^{-3/2\beta},
\end{equation}
while the ICM temperature is constant. Here we
 present simulations for three clusters. The parameters are listed in Table~\ref{tab:ICM}.
 \begin{table}
 \caption{ICM parameters: Core radius, $R\ICM$, beta-parameter, $\beta$,
   central ICM density, $\rho\ICM{}_0$, and ICM temperature, $T\ICM$, for
   clusters C1, C2 and C3.}
 \label{tab:ICM}
 \centering\begin{tabular}{lcccc}
 \hline
     &$R\ICM/\Kpc$  & $\beta$ & $\rho\ICM{}_0/(\gccm)$    & $T\ICM/\K$ \\
 \hline
 C1  & 50      & $0.5$   & $2\cdot 10^{-26}$ & $4.7\cdot 10^7$ \\
 C2 & $\cdot$ & $\cdot$  & $10^{-26}$        & $\cdot$ \\
 C3 & 386     & 0.705    &  $6.07\cdot 10^{-27}$  & $9.5\cdot 10^7$\\
 \hline
 \end{tabular}
 \end{table}
%
Figure~\ref{fig:cluster_profiles} shows the density and pressure profiles of
clusters C1 and C3.
%
%
 The only difference between clusters C1 and C2 is a factor of 2 in the
 density. These two clusters are compact, the ICM density is strongly peaked.
 Thus, they are similar to the Virgo cluster (\citealt{matsumoto00}), but not
 as extreme as Virgo (see Fig.~\ref{fig:cluster_profiles}).  Cluster C3 resembles the Coma cluster (parameters see
 \citealt{mohr99}), which is very extended.

\subsection{Galaxy orbit} \label{sec:orbits}
The orbit of the galaxy determines its ram pressure history. According to
the simulations of RB06\nocite{roediger06}, for our model galaxy, complete
stripping is expected for ram pressures of about $10^{-10}\Erg\ccm$. We will
refer to such ram pressures as ``high''. ``Medium'' ram pressures ($\sim
10^{-11}\Erg\ccm$) are still expected to strip the gas disc significantly.  We
aimed at constructing orbits with medium to high ram pressure peaks.
Moreover, we concentrate on orbits of galaxies that could be regarded as
falling into the cluster for the first time, i.e. they start from a
sufficiently large distance from the cluster centre. Additionally, the
galaxies are bound to the cluster.

\begin{figure}
 \centering\resizebox{\hsize}{!}{\includegraphics[angle=-90]{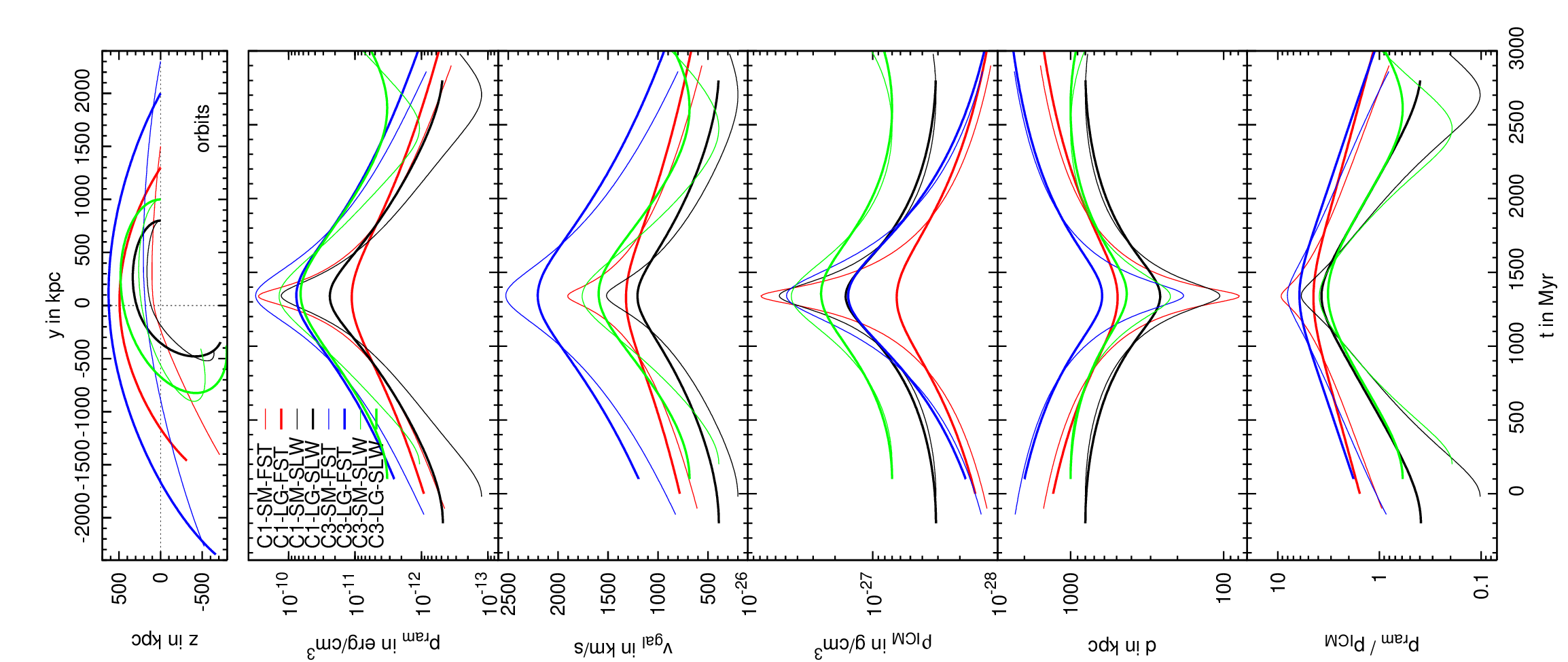}}
 \caption{Representative galaxy orbits in clusters C1 and C3: The top panel shows the orbits, which are in the
   $y$-$z$-plane. The next panels show the temporal evolution of  ram pressure,
   $p\Ram$, galaxy velocity, $v\Gal$, ICM density along orbit, $\rho\ICM$,
   distance to cluster centre, $d$, and ratio between ram pressure and local
   ICM pressure, $p\Ram/p\ICM$. The clusters (C1 or C3), typical galaxy velocity (FaST or
   SLoW) and
   impact parameter (SMall or LarGe) are coded by line colour and thickness, see legend.}
 \label{fig:orbits_profiles_more}
 \end{figure}

\begin{figure}
 \centering\resizebox{\hsize}{!}{\includegraphics[angle=-90]{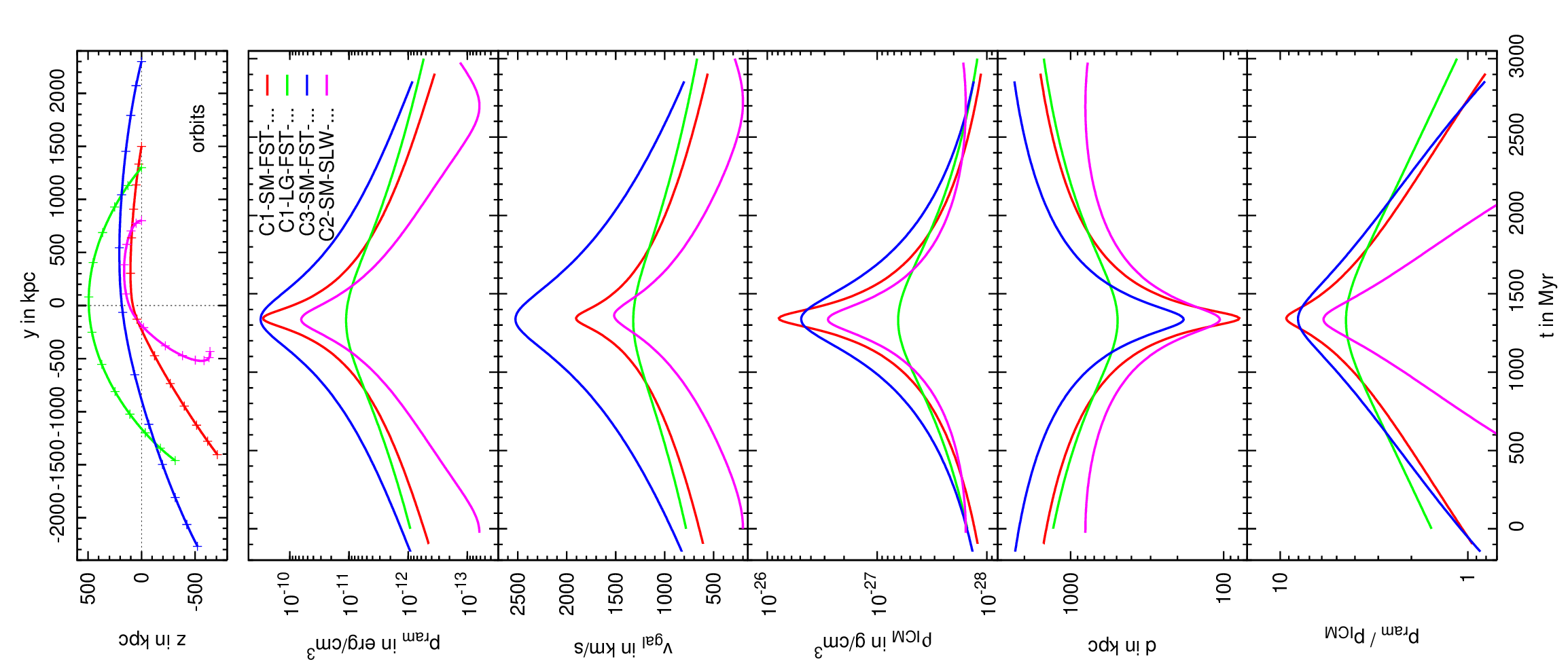}}
 \caption{Summary of galaxy orbits for our simulations. The top panel shows the orbits, which are in the
   $y$-$z$-plane. Crosses mark the position of the galaxy in intervals of
   $250\Myr$. The next panels show the temporal evolution of  ram pressure,
   $p\Ram$, galaxy velocity, $v\Gal$, ICM density along orbit, $\rho\ICM$,
   distance to cluster centre, $d$, and ratio between ram pressure and local
   ICM pressure, $p\Ram/p\ICM$. For an explanation of the labels see text.}
 \label{fig:orbits_profiles}
 \end{figure}

Figure~\ref{fig:orbits_profiles_more} shows several representative orbits in
clusters C1 and C3. 
%
%
Already at this stage some important issues can be learned:
\begin{itemize}
\item Naturally, orbits with large impact parameters yield ram pressure peaks
  of long duration. Short ram pressure peaks are typical for galaxies passing
  near the cluster centre. However, even on orbits with small impact
  parameters, the ram pressure peaks in the extended cluster are about twice
  as wide as on comparable orbits in the compact cluster. 
\item Orbits with low ram pressure peaks are practically impossible in both,
  the extended and the compact cluster. Even on orbits with large impact
  parameter, where the ICM density along the orbit is small, the galaxy's
  velocity is high enough to yield at least medium ram pressures. If the
  galaxy passes near the cluster centre, typically both, ICM density and
  galaxy velocity are high.
\item In the extended cluster, even orbits with medium ram pressure peaks are
  difficult to construct. Almost all galaxies should be stripped severely or
  completely.
\item Comparing the pressure profile along the galactic disc as shown in
  Fig.~\ref{fig:initial_profiles} with the radial pressure profiles of the
  clusters (Fig.~\ref{fig:cluster_profiles}) gives an interesting insight into
  the importance of the cluster environment for the galaxy: The ICM-pressure
  in the Coma-like cluster drops below $10^{-12}\Erg\ccm$ only for radii
  larger than $\sim 2\Mpc$. Inside $\sim 1\Mpc$ the ICM pressure is above
  $10^{-11}\Erg\ccm$. These values are comparable to the pressure in the
  galactic plane at disc radii of approx.~15 and $8\Kpc$. Thus, even if a
  galaxy could reach such a position in the cluster without suffering from any
  other processes, its gas disc should be affected at least by the external pressure.
\item During most of the orbit, the ram pressure is even larger than the thermal ICM
  pressure (bottom panel of Fig.~\ref{fig:orbits_profiles}).
\end{itemize}

Here we present simulations for four orbits that are summarised in
Fig.~\ref{fig:orbits_profiles}.
%
%
The labels of the runs represent the cluster (C+number), small or large impact
  parameter (SM or LG), fast or slow galaxy (FST or SLW), inclination (F for
  near face-on, M for medium, E for near edge-on; FE for first near face-on
  but then near edge-on, etc.).
 In all cases, the galaxy orbits in the $y$-$z$-plane. 
Table~\ref{tab:runs} lists the simulation runs.
%
\begin{table}
\caption{List of runs. Run labels code the cluster (C+number), small or large impact
  parameter (SM or LG), fast or slow galaxy (FST or SLW), inclination (F for
  near face-on, M for medium, E for near edge-on; FE for first near face-on
  but then near edge-on, etc.). Initial galaxy coordinates $x\Gal{}_0$ and
  $z\Gal{}_0$ are always zero, $y\Gal{}_0$ is given in the second
  column. The third column lists the initial galaxy velocity. The inclination
  listed in the fourth
  column is the angle between the galaxy's rotation axis and the $y$-axis.}
\label{tab:runs}
\centering\begin{tabular}{lrrr}
\hline
label     & $y\Gal{}_0$ & initial velocity & inclination  \\
          &  in kpc             &  in $\Kms$       & in $\degree$ \\
\hline
C1-SM-FST-MF  &  1500 & $(0, -600,100)$ & $-30$ \\
C1-SM-FST-E  &  $\cdot$      &     $\cdot$     & $80$ \\
\hline
C1-LG-FST-MF &  1300 & $(0, -600,500)$ & $-30$ \\
C1-LG-FST-EF &  $\cdot$      & $\cdot$         & $-60$ \\
C1-LG-FST-FE &  $\cdot$      & $\cdot$         & $45$ \\
\hline
C2-SM-SLW-FME &  800  & $(0, 0,200)$  & $30$ \\
C2-SM-SLW-EMF &  $\cdot$  & $\cdot$  & $-60$ \\
C2-SM-SLW-MFM &  $\cdot$  & $\cdot$  & $-20$ \\
\hline
C3-SM-FST-MF  & 2300 & (0,-800,200) & 30 \\
\hline
\end{tabular}
\end{table}
%


\section{Results}
%
\label{sec:results}

Previous simulations where the galaxy was exposed to a constant ICM wind found
that the ICM-ISM interaction is a multi-stage process (SS01, MBD03, RH05, RB06\nocite{schulz01,marcolini03,roediger05,roediger06}). The stages reflect the
two gas removal mechanisms that work on different timescales.

One process is ram pressure ``pushing'', the immediate consequence of the ram
pressure dislocating the gas disc at radii where the galaxy's gravity is not
strong enough. This process results in the instantaneous stripping phase where
the gas is pushed out of its original position on a timescale of a few
$10\Myr$. However, the gas does not become unbound from the galactic potential
immediately but with a certain delay, because it takes some time until the gas
is accelerated to the escape velocity. This time
delay grows with decreasing ram pressure as the ram pressure is responsible
for accelerating stripped gas packages.

The second gas removal process is continuous stripping (or turbulent/viscous
stripping, see e.g.~\citealt{nulsen82,quilis00}), which is caused by the
Kelvin-Helmholtz instability induced by the ICM wind flowing over the
surface of the gas disc. Continuous stripping leads to a slow but
continuous gas loss at a rate of $\sim 1M\Sun\Yr^{-1}$, it works on a
timescale of Gyrs. The gas disc radius is not changed significantly
by continuous stripping.

\subsection{Analytical estimates}
\label{sec:analytical}
The usual way to estimate the amount of gas loss due to ram pressure pushing
for galaxies moving face-on follows the suggestion of GG72. Here,
one compares the gravitational restoring force per unit area,
%
\begin{equation}
f\Grav(R) = \Sigma\Gas(R)\; \frac{\partial\Phi}{\partial Z}(R),   \label{eq:fgrav}
\end{equation} 
%
 and the ram
pressure,
%
\begin{equation}
p\Ram = \rho\ICM v\Gal^2, \label{eq:pram}
\end{equation} 
%
 for each radius of the galaxy (see e.g.~RB06\nocite{roediger06}), where $\Phi$ is the gravitational potential of the galaxy, $\Sigma\Gas(R)$
the gas surface density, $\rho\ICM$ the ICM density and $v\Gal$ the galaxy's velocity.  At
radii where the restoring force is larger, the gas can be retained, at radii
where the ram pressure is larger, the gas will be stripped. The transition
radius is called the stripping radius.
In order to apply this estimate to our simulations, we compare the
gravitational restoring force to the current ram pressure at every time. This estimate
should work best for galaxies moving face-on. 

(Nearly) pure continuous stripping affects galaxies moving edge-on. All
galaxies with other inclinations experience a mixture of ram pressure pushing
and continuous stripping.  For a spherical gas cloud, the mass loss rate due
to continuous stripping was estimated by \citet{nulsen82}. For a sphere of
radius, $R$, moving though the ICM of density, $\rho\ICM$, with velocity,
$v\Gal$, the appropriate mass loss rate is $\pi R^2\rho\ICM v\Gal$.  As the
surface area of a sphere is twice as large as the surface area of a disc
(counting upper and lower side), we use
%
\begin{equation}
\dot M\KH = 0.5\pi R^2 \rho\ICM v\Gal \label{eq:KH}
\end{equation} 
%
for our case of a disc galaxy. Both, ICM density, $\rho\ICM$, and galaxy
velocity, $v\Gal$ are time-dependent. As disc radius, $R$, we use the current
analytical stripping radius as estimated by the GG72 criterion, which is
also time-dependent. This choice may seem mismatched at first, but it is
motivated by the fact that the GG72 criterion gives good results for the
stripping radius for most inclinations.

\subsection{Snapshots}
\begin{figure*}
\includegraphics[angle=0,width=0.45\textwidth]{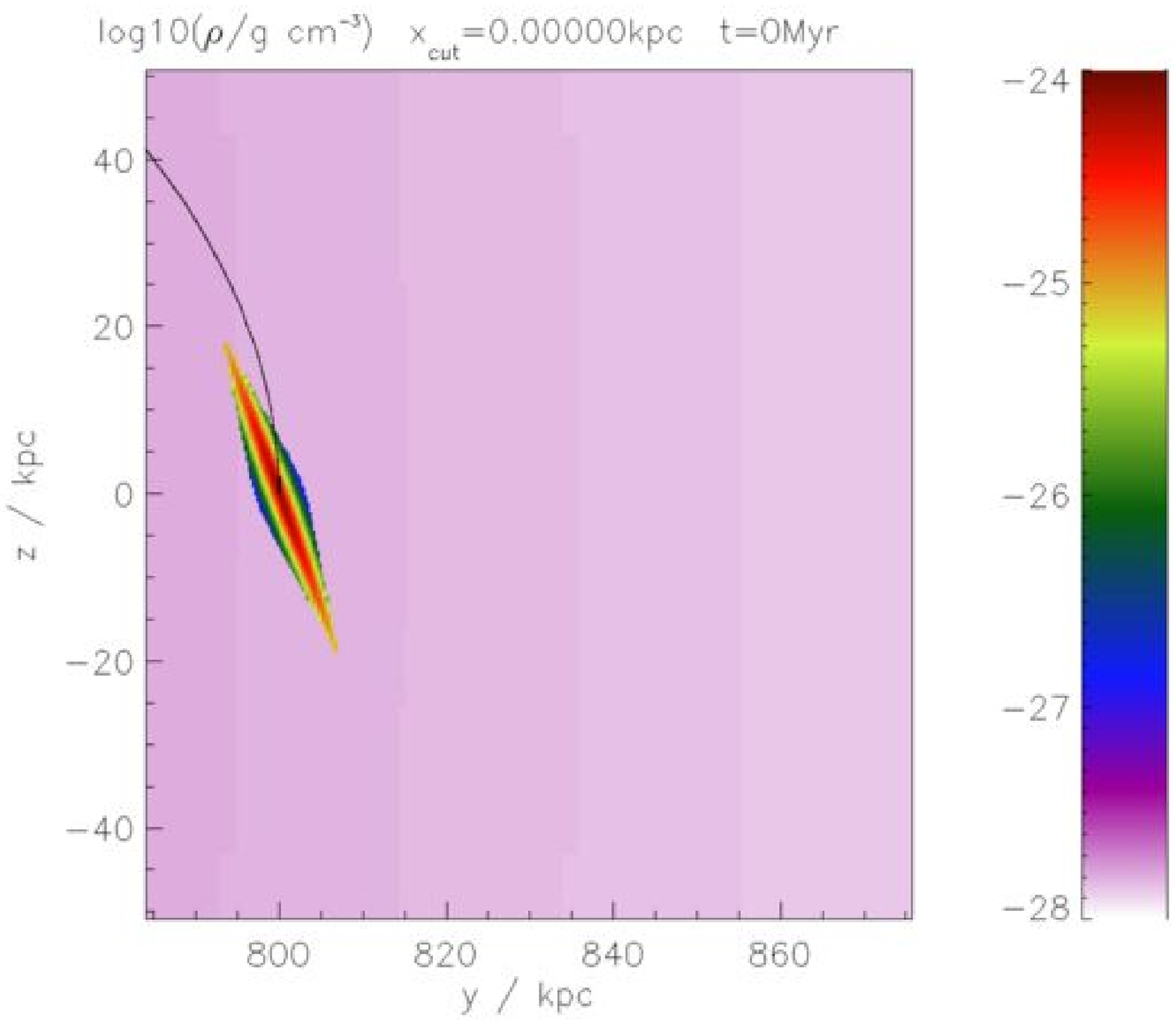}
\includegraphics[angle=0,width=0.45\textwidth]{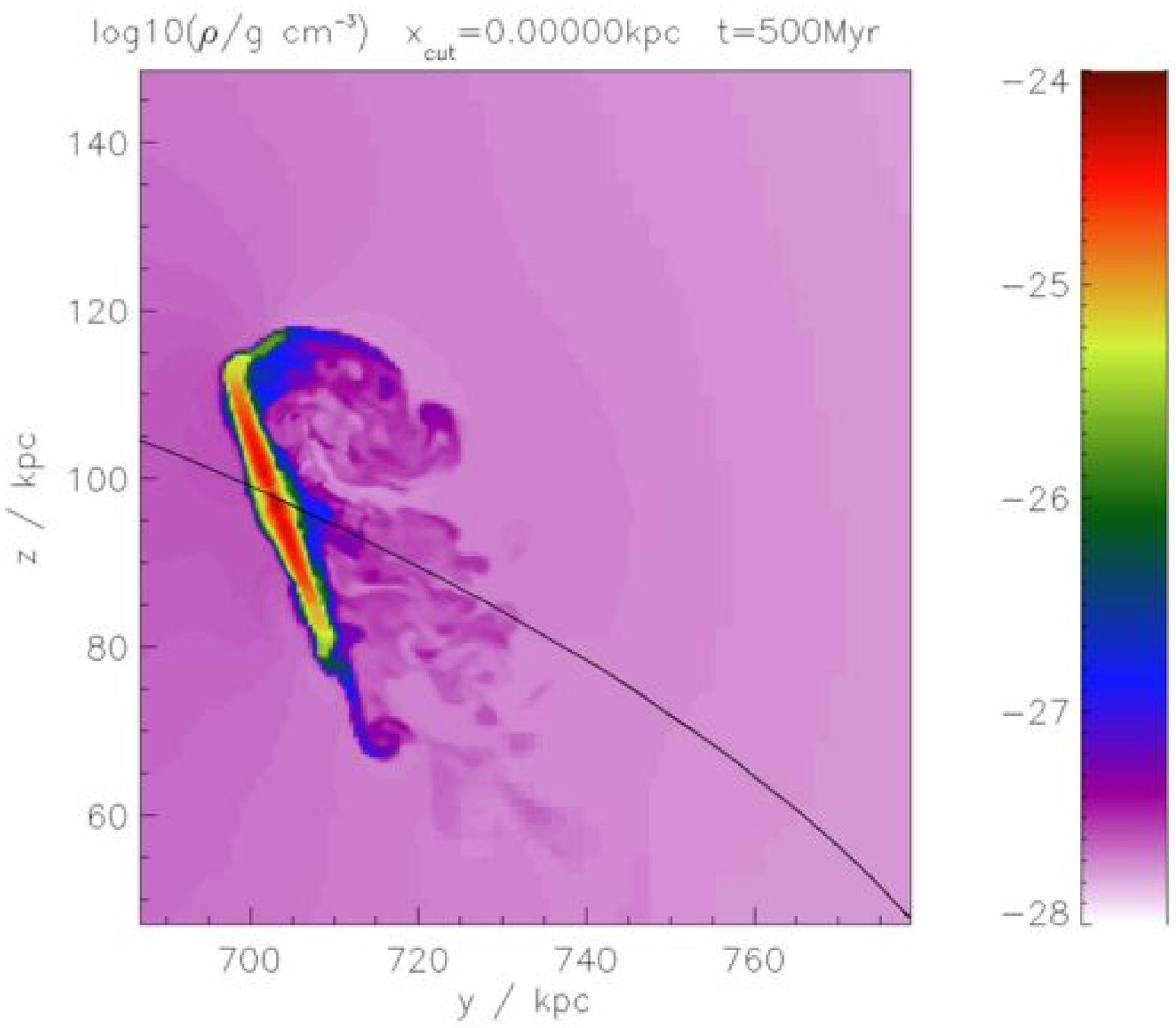}
\includegraphics[angle=0,width=0.45\textwidth]{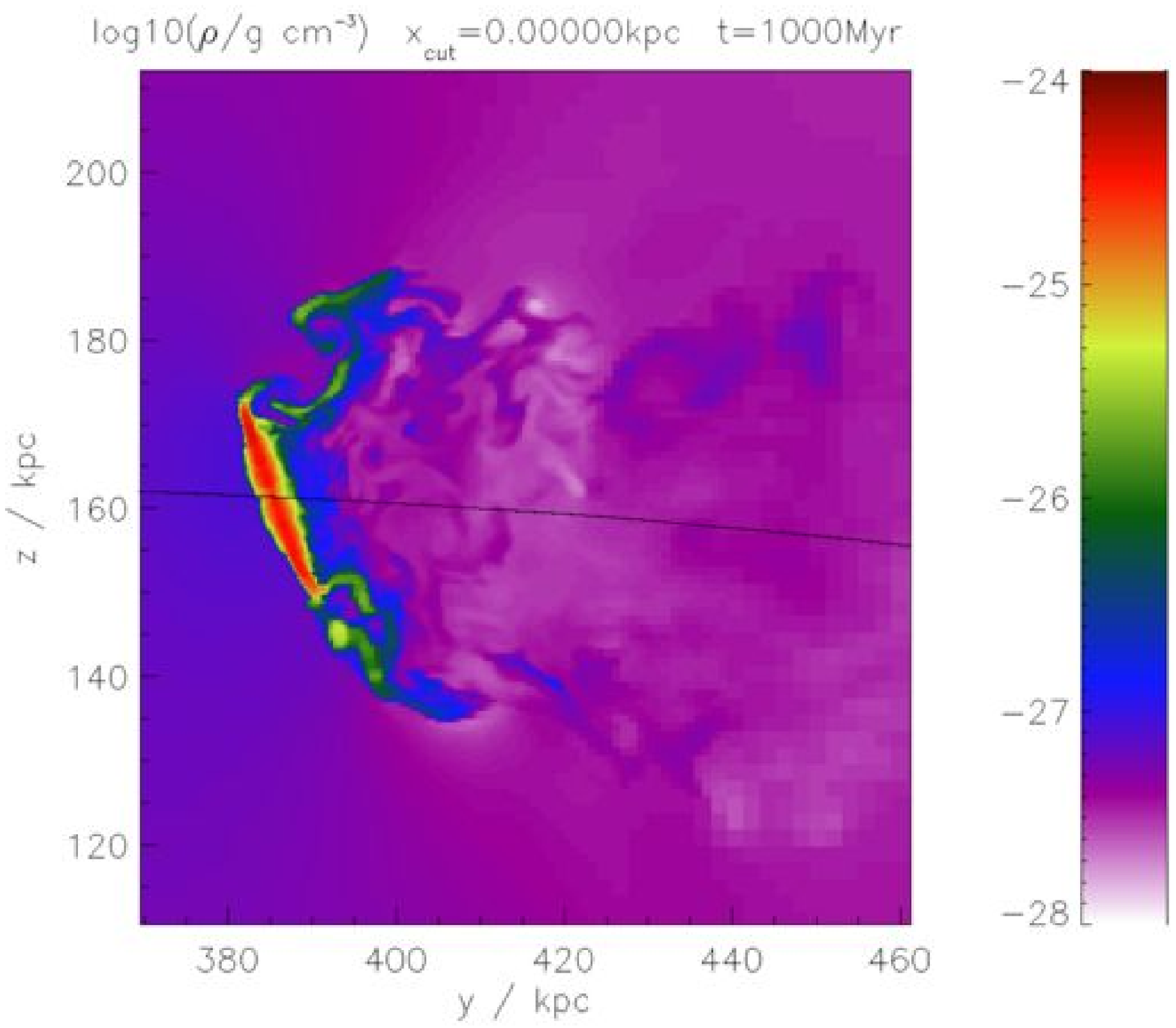}
\includegraphics[angle=0,width=0.45\textwidth]{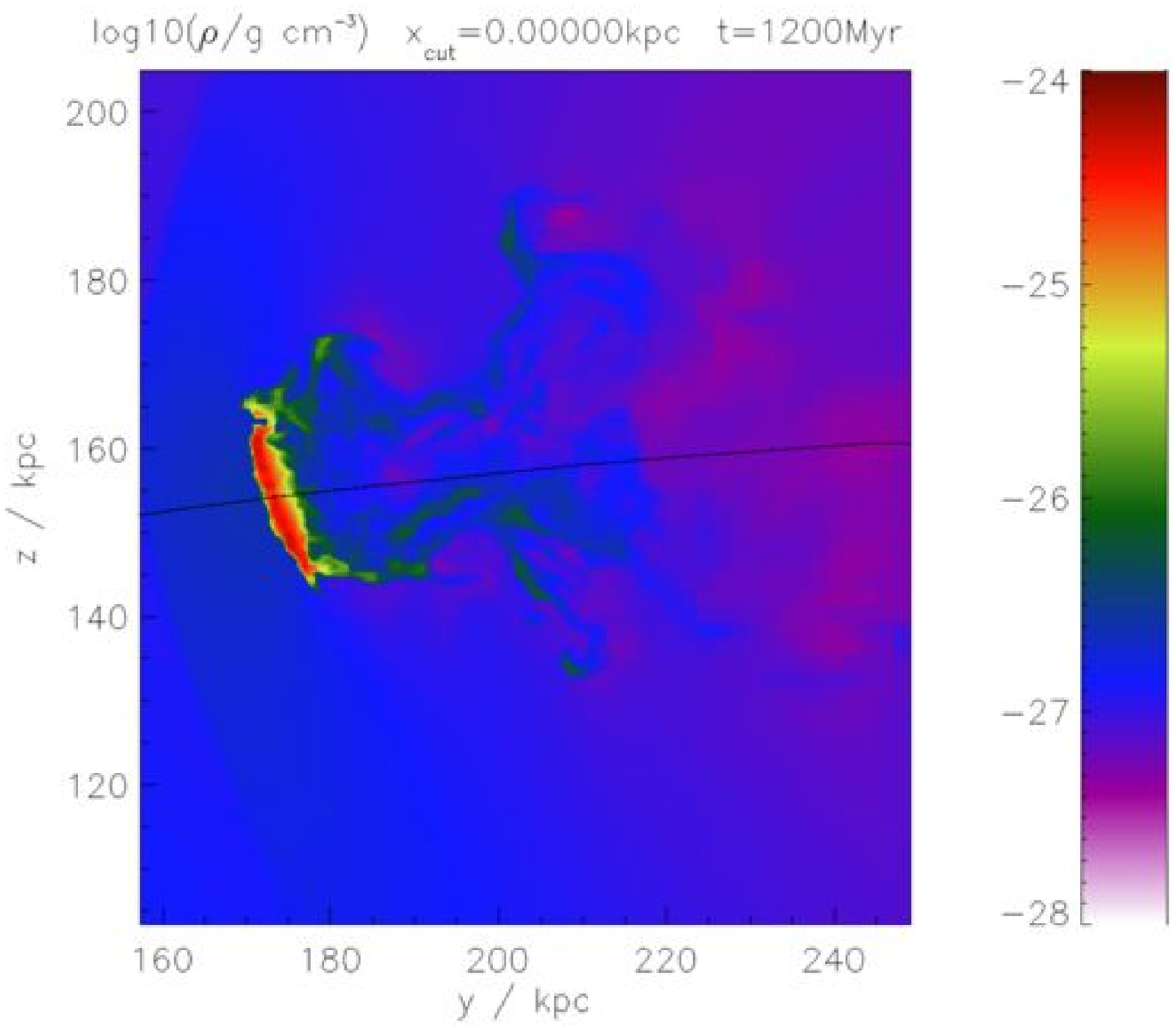}
\includegraphics[angle=0,width=0.45\textwidth]{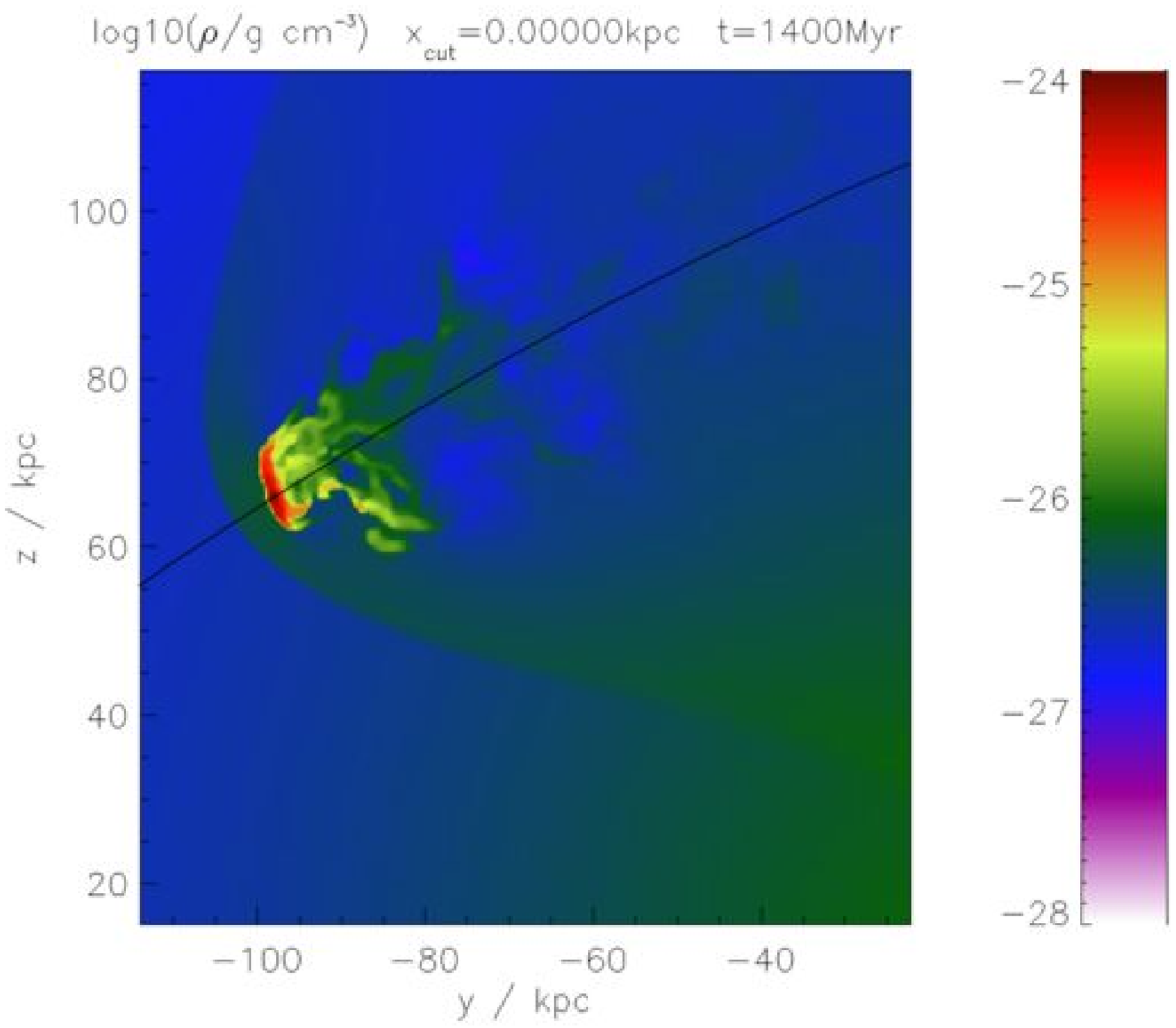}
\includegraphics[angle=0,width=0.45\textwidth]{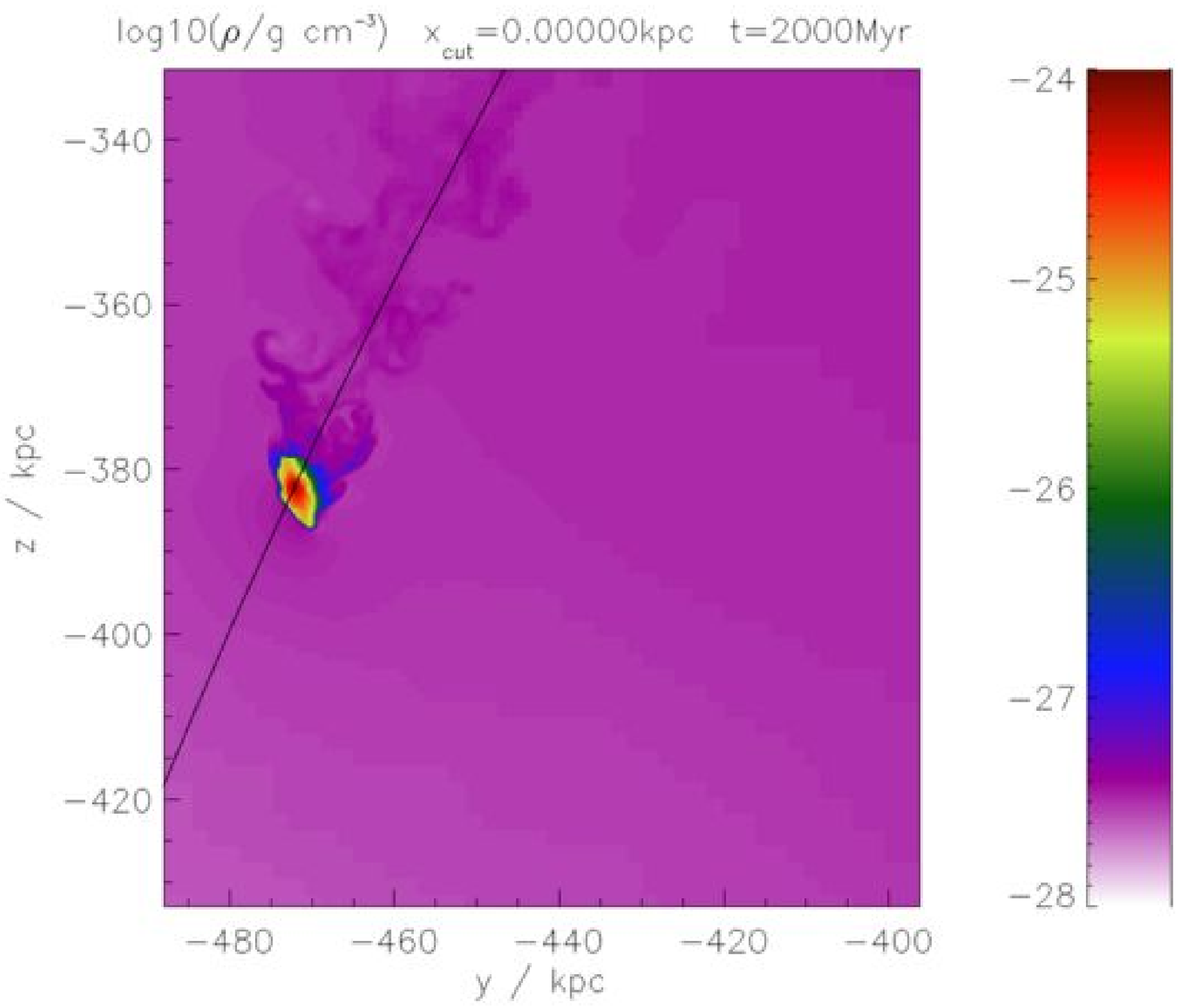}
\caption{Snapshots for run C2-SM-SLW-MFM (high resolution): colour-coded gas
  density in the orbital plane. The black line marks the galaxy orbit.}
\label{fig:snapshots}
\end{figure*}

Figure~\ref{fig:snapshots} shows snapshots for run C2-SM-SLW-MFM (high
resolution).
%
%
At larger distances to the cluster centre, the galaxy moves subsonically, in the inner
cluster it moves supersonically as indicated by the bow shock. Before pericentre
passage, the gas disc size decreases, whereas it stays rather constant after
pericentre passage. Snapshots for other runs are similar.

\subsection{Comparison with numerical result}
\label{sec:comp_num_an}
\begin{figure*}
\includegraphics[angle=-90,width=0.45\textwidth]{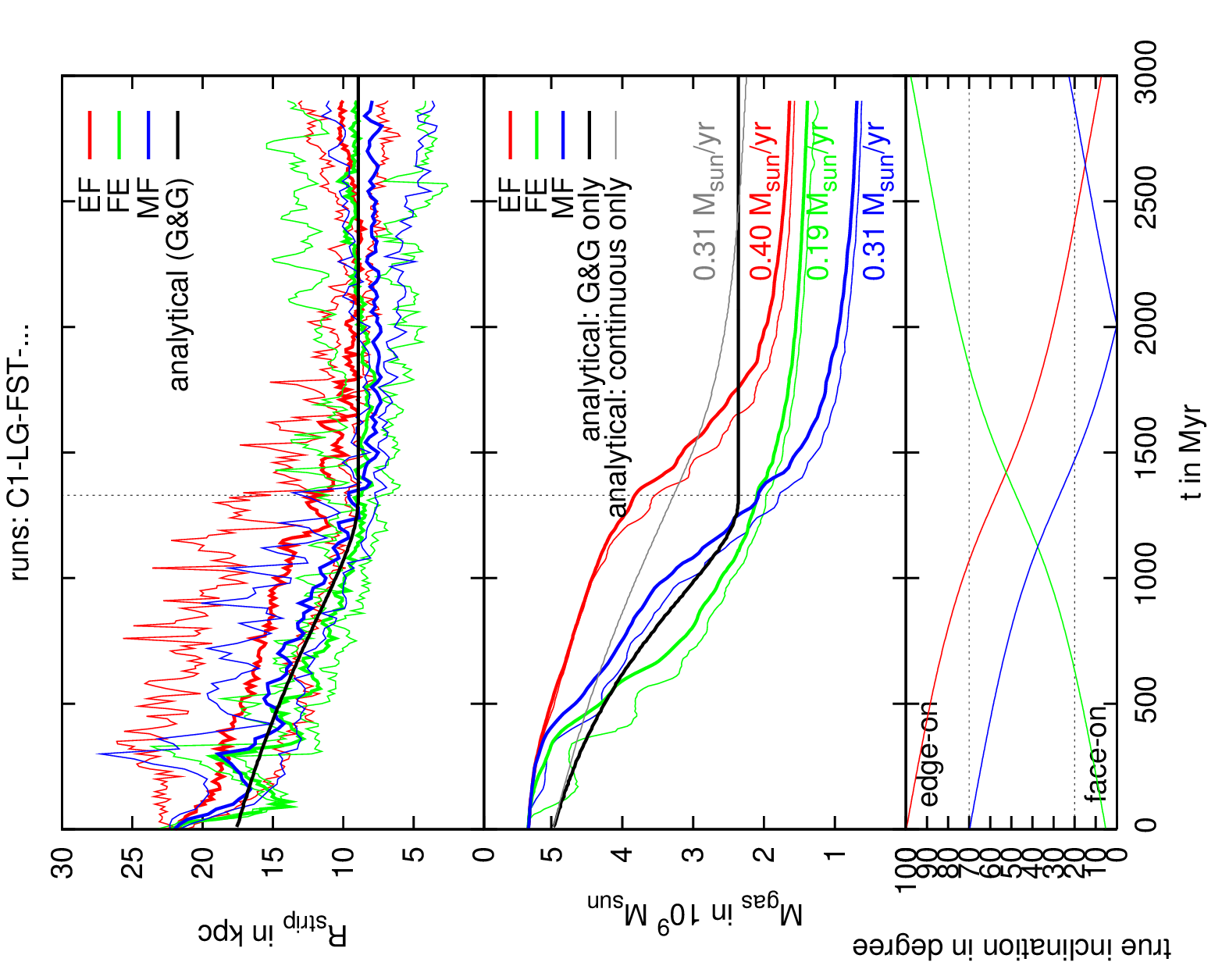}
\includegraphics[angle=-90,width=0.45\textwidth]{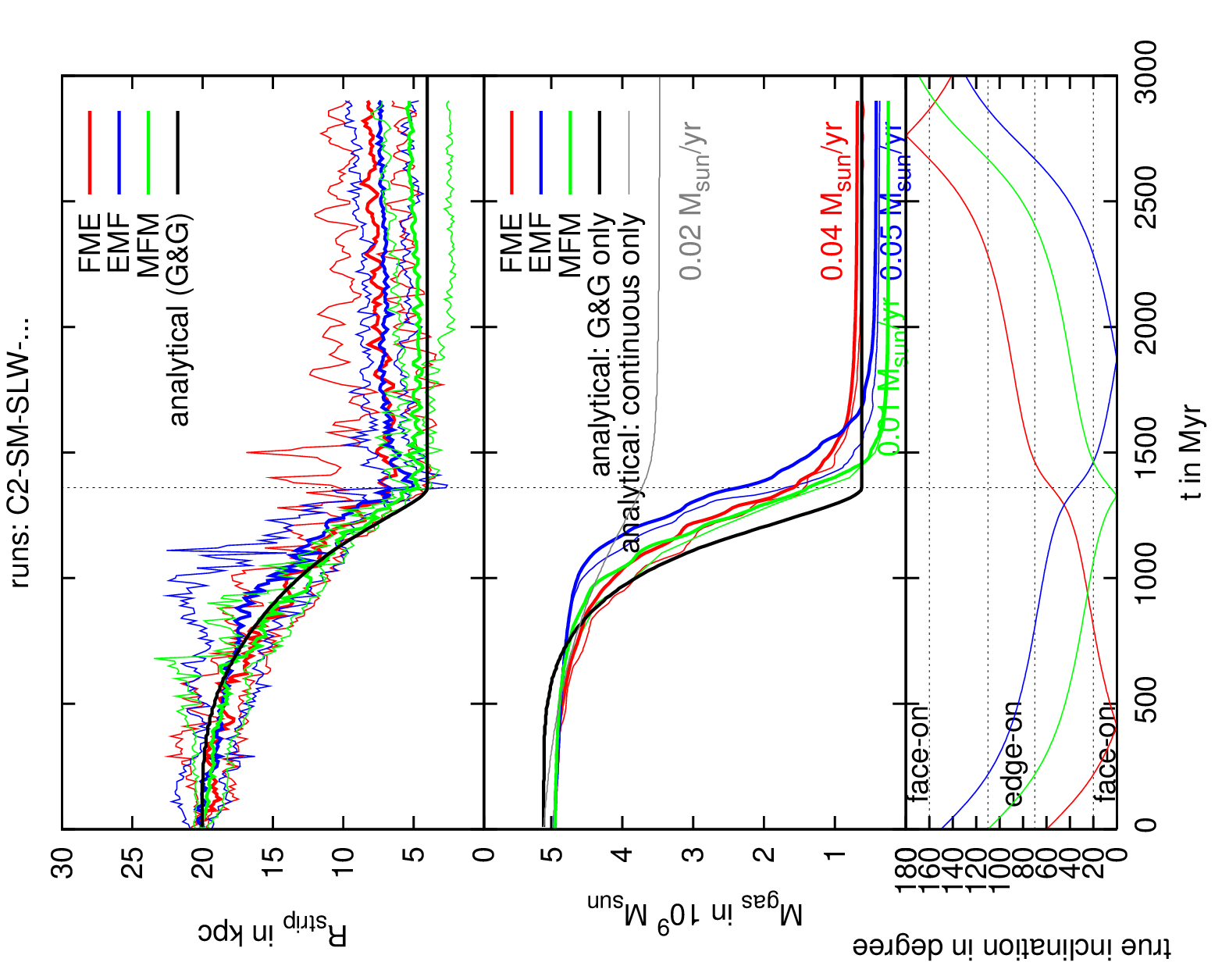}
\includegraphics[angle=-90,width=0.45\textwidth]{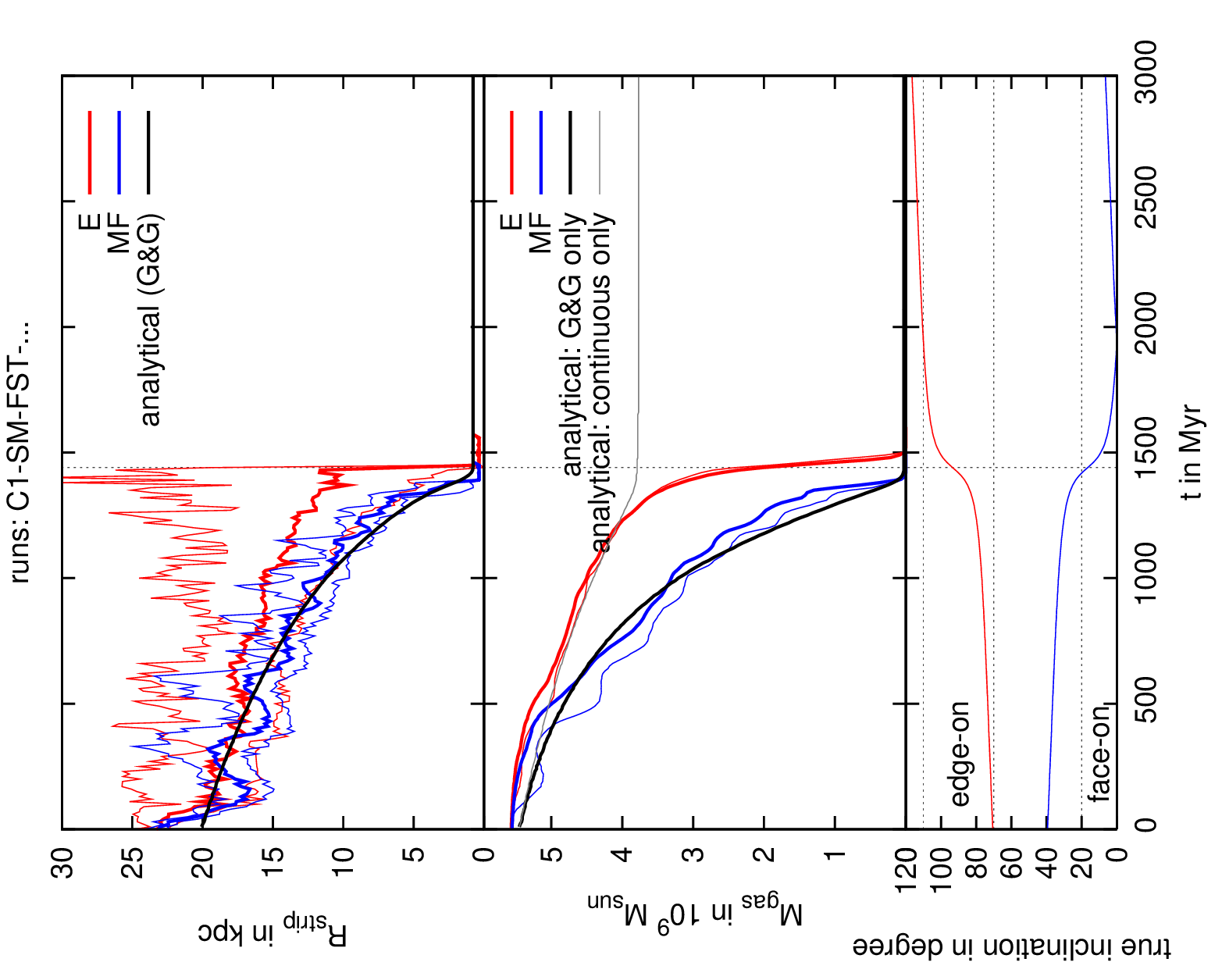}
\includegraphics[angle=-90,width=0.45\textwidth]{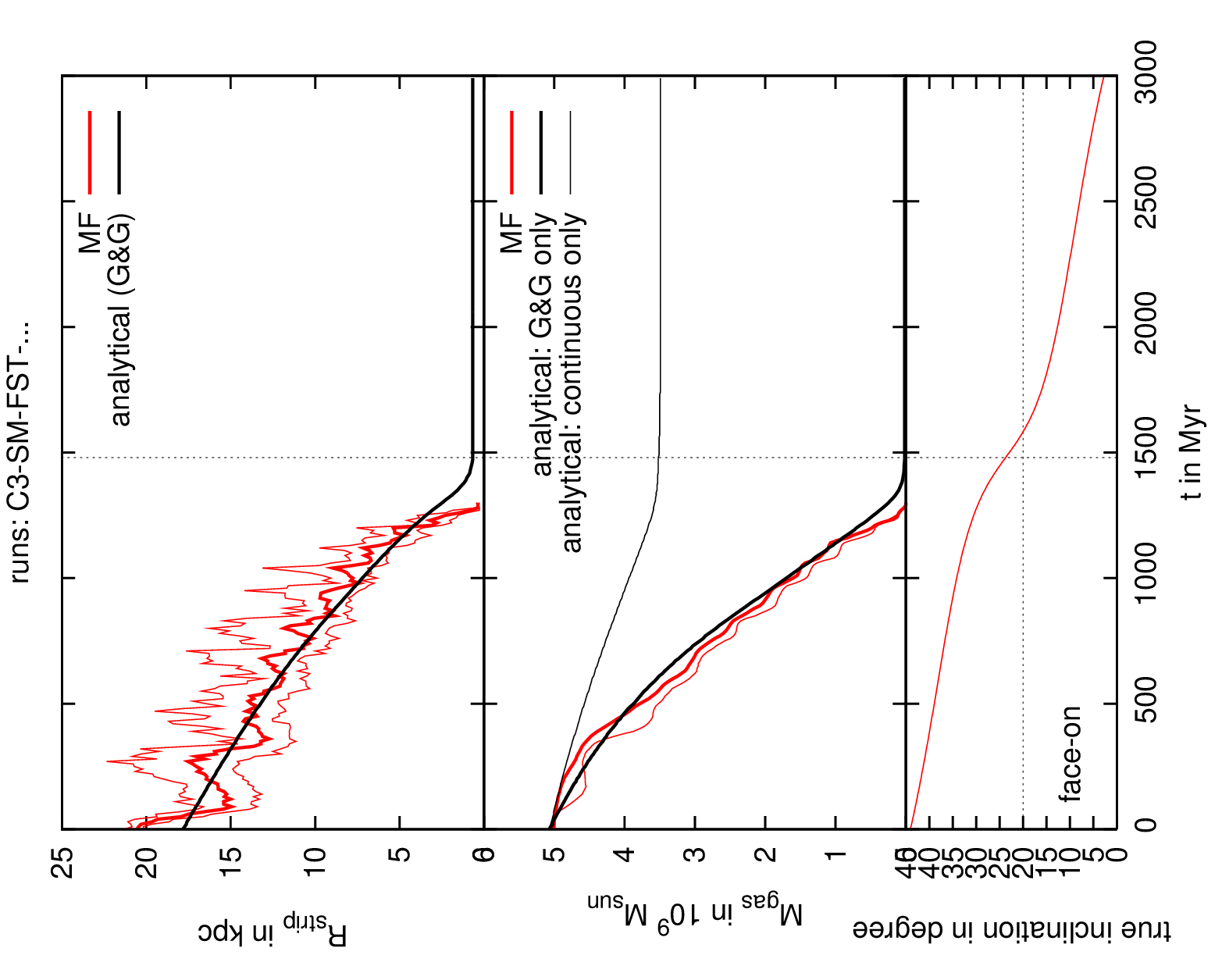}
\caption{Comparison between analytical and numerical stripping radius (top panels) and
  gas mass (middle panels).  The bottom panels display the evolution of
  the true inclination, i.e. the angle between the galaxy's rotation axis and
  the direction of motion.  Each subfigure is for one orbit (see title),
  different inclinations are colour-coded (see legend). For the numerical
  stripping radius, the mean value (thick lines) as well as maximum and
  minimum radius (thin lines) as described in RB06 are shown.
  For the gas mass, we show the bound gas mass (thick lines) and the mass of
  gas in the disc region (thin lines of matching colour, see also RB06).
  Analytical estimates according to the Gunn\&Gott criterion are shown as
  black lines for the stripping radius as well as for the remaining gas mass.
  The gray line in the mass plots displays the predicted remaining gas mass if
  only continuous stripping is taken into account. In cases where the galaxy
  was not stripped completely, we did a linear fit for the bound gas mass
  for the final Gyr of the simulations as well as for the analytical estimate
  of continuous stripping. The resulting mass loss rates are denoted
  colour-coded in the plot.}
\label{fig:comp_an_num}
\end{figure*}

Figure~\ref{fig:comp_an_num} compares analytical estimates and numerical
results for the stripping radius and the remaining gas mass.
%
%
On the way towards the cluster centre, the distinction between the instantaneous
and continuous stripping phases vanishes as both processes overlap. In earlier
simulations with constant ICM wind (RH05, RB06, but also
MBD03\nocite{roediger05,roediger06,marcolini03}) a short period of gas
backfall during the intermediate phase was observed. The gas disc mass as a
function of time showed a characteristic local maximum after the instantaneous
stripping phase. Now, in the simulations with varying ICM wind, we observe a
repeated backfall of stripped gas prior and during
pericentre passage. The backfall appears as repeated local maxima or plateaus in the
gas disc mass as a function of time.  Interestingly, the sticky-particle
simulations of \citet{vollmer01a} where the galaxies were also exposed to a varying
ram pressure found gas backfall of up to $5\times 10^8 M\Sun$ {\em after}
pericentre passage, primarily for edge-on galaxies. This behaviour cannot be
observed in our simulations. The backfall cycles
in our simulations are caused by the hydrodynamical velocity field in the
galaxy's wake.  On the way out of the cluster, the instantaneous stripping is
not active anymore because all gas from outer radii is already lost. Here
continuous stripping dominates.

\subsubsection{Ram pressure pushing}
The result of the analytical estimate that only takes ram pressure pushing into
account is shown as black lines in Fig.~\ref{fig:comp_an_num}. The analytical
and numerical result for the stripping radius agree remarkably well. Only
during the pericentre passage in runs C2-SM-SLW-\ldots, the simulated galaxies
can retain a larger gas disc than predicted.  Here, the ram pressure, $p\Ram$,
increases faster than the stripping timescale: During the $\sim 200\Myr$
before the ram pressure maximum, the stripping radius is predicted to decrease
from $10\Kpc$ to $4\Kpc$, which corresponds to an average radius decrease of
$1\Kpc$ per $35\Myr$. A rough estimate of the acceleration of
 gas with a surface density, $\Sigma$, is $p\Ram/\Sigma$.
Thus, the stripping timescale, i.e~the time needed to move the gas package a
distance, $s$, is
%
\begin{equation}
\tau=55\Myr\;  \sqrt{  \frac{s}{5\Kpc} 
             \;   \frac{\Sigma}{10^{-3}\mathrm{g}\,\Cm^{-2}}  
             \;   \frac{10^{-11}\Presunit}{p\Ram}  },
\end{equation} 
%
where we have inserted typical values for this case. However, this estimate is a lower
limit, as it neglects the effect of the galaxy's restoring force, which
lessens the effective acceleration. Although this timescale is rather short,
in runs C2-SM-SLW-\ldots the ram pressure changes on an even faster timescale
during pericentre passage.

For all other runs, deviations of the gas disc radius from the analytical result 
can be attributed to different inclinations and they agree with the results of
constant ICM wind simulations of RB06\nocite{roediger06}. As expected from the
simulations of RB06, with a increasing deviation from the face-on inclination,
the remaining gas disc becomes increasingly asymmetrical. This behaviour
translates into an increasing distance between the minimum and maximum radius
lines in Fig.~\ref{fig:comp_an_num}.  A further agreement with the constant
wind simulations is the fact that for the highest ram pressure peak, the
galaxy is stripped completely even if it is moving near-edge-on (run C1-SM-FST-E).

The slightly premature complete stripping of the galaxies in runs C1-SM-FST-MF
and C3-SM-FST-MF is caused by the fact that the remaining gas disc is affected
by numerical diffusion due to the relatively low resolution of $0.5\Kpc$.
Details are discussed in App.~\ref{sec:resolution}.  All other results
concerning ram pressure pushing are not affected by numerical diffusion or
resolution.

For the remaining gas mass, the simple analytical estimate and the numerical
results agree only roughly. Generally, in the simulations the galaxies lose
their gas more slowly than predicted, because the analytical estimate does not
include the delay in mass loss explained above
(Sect.~\ref{sec:analytical}). Moreover, it neglects effects of inclination and
continuous stripping.

\subsubsection{Continuous stripping}

The gray line in the mass plots in Fig.~\ref{fig:comp_an_num}
displays the analytical estimate for the remaining gas mass if purely continuous
stripping (see Eq.~\ref{eq:KH}) is taken into account. Before pericentre
passage, this is most
likely to apply to galaxies moving edge-on. Indeed, we find a
good agreement in mass loss rate for the first Gyr of runs C1-LG-FST-EF and C1-SM-FST-E, where
this condition is fulfilled. For  C2-SM-SLW-EMF, the ram pressure is very
small during the first Gyr and thus the delay in mass loss is long. Therefore,
the mass loss rate is overpredicted here.

Pure continuous stripping is also expected to apply to the time well after
pericentre passage. Thus, we have calculated averaged mass loss rates for the
final Gyr of the simulations.  We have also calculated the averaged mass loss
rate for this time interval from the analytical prediction by
Eq.~\ref{eq:KH}. The values are given in Fig.~\ref{fig:comp_an_num}. 
The mass loss rates at this late stage of the simulations have to be
interpreted as upper limits, because due to the relatively low resolution of
$0.5\Kpc$ the remaining gas disc is already  significantly affected by numerical diffusion
(see Sect.~\ref{sec:resolution} for a detailed discussion).


\section{Discussion and summary}
%
We have presented 3D hydrodynamical simulations of RPS of a disc galaxy
orbiting in a galaxy cluster. We have focused on the evolution of the radius
of the remaining gas disc and the mass of gas bound to the galaxy. In
particular, we compared the numerical results to a time-dependent version of
the classical GG72 estimate.

For galaxies not moving near edge-on, we find that the stripping radius is
predicted well by the estimate based on the GG72 criterion, unless the
ram pressure increases faster than the stripping timescale. Then the remaining
gas disc is somewhat larger than predicted. However, the evolution of the
bound gas mass differs between simulations and the
analytical estimate: In the simulations, the galaxy loses the gas more slowly than
predicted. Deviations of the analytical estimate from the simulation results are caused by
effects that are not taken into account by the estimate: the inclination of the galaxy, the
time delay of true gas loss, and the combined effect of ram pressure pushing
and continuous stripping. Although our simulations are an important step
towards realistic models of RPS, they also neglect some aspects that have to be studied in
the future, e.g. the multiphase physics of the ISM and transport processes
and ambient motions in the ICM.

The differences between analytical estimate and simulations in the evolution
of the remaining gas mass will also be reflected in the distribution of
stripped ISM throughout the cluster. If galaxies followed the prediction of
the estimate based on the GG72 criterion, they would not lose gas after
pericentre passage. In the simulations, however, they do. Thus, the
distribution of the stripped ISM along the orbit is broader than expected.



\section*{Acknowledgements}
We acknowledge the support by the DFG grant BR 2026/3 and the supercomputing grants NIC
2195 and 2256 at the John-Neumann Institut (NIC) in J\"ulich.
The results presented were produced using the FLASH code, a product of the DOE
ASC/Alliances-funded Center for Astrophysical Thermonuclear Flashes at the
University of Chicago.

\appendix
\section{Influence of resolution}
%
\label{sec:resolution}
In order to check the influence of the resolution on our results, we repeated
two runs with better resolution. Run C1-LG-FST-MF was repeated such that the
resolution was improved by a factor of 2 everywhere in the grid compared to
the description in Sect.~\ref{sec:code}. Run C2-SM-SLW-MFM was repeated such
that the inner disc region, i.e. a cylinder around the galactic centre with
radius $10\Kpc$ and thickness $6\Kpc$, was always refined to $0.25\Kpc$.

\begin{figure}
 \centering\resizebox{\hsize}{!}{\includegraphics[angle=0]{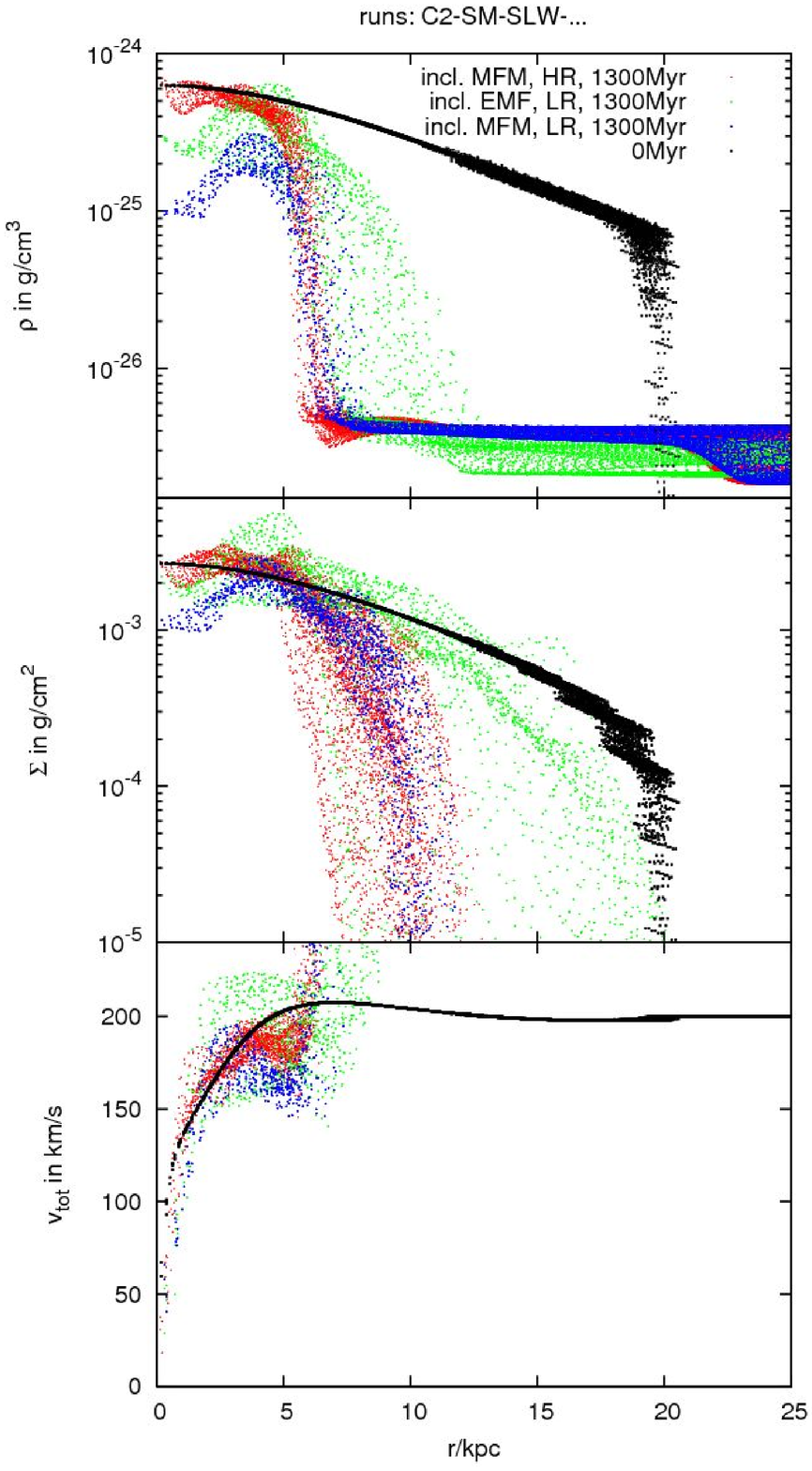}}
 \caption{Radial profiles of the density, $\rho$, projected surface density, $\Sigma$,
   and total velocity, $v_{\mathrm{tot}}$, in the galactic plane at
   $t=1300\Myr$ for different runs along the orbit C2-SM-SLW (see legend). 
   One dot for each cell in the galactic plane is plotted. Black dots show
   the initial profile. Red and blue dots are for runs with
   inclination MFM, but different resolution (red for $\Delta x=0.25\Kpc$,
   denoted ``HR'' in the legend; blue for $\Delta x=0.5\Kpc$, denoted
   ``LR''). Green dots are for inclination EMF and also resolution $\Delta x=0.5\Kpc$.}
 \label{fig:comp_prof}
 \end{figure}

In grid-codes, advection of mass and momentum
into non-axial directions is always accompanied by a numerical viscosity due
to numerical diffusion. In the rotating gas disc, the gas has to move on
circular orbits in a Cartesian grid. This geometrical mismatch is strongest in
the inner part of the galaxy. Below a certain radius, the resolution of the
rotational motion becomes insufficient and the numerical diffusion introduces
radial velocities, which lead to a decrease of the central density.
Fig.~\ref{fig:comp_prof} shows an example for this effect.
%
%
As expected, the decrease in central density is much weaker if if higher
resolution is used. Interestingly, it is also less pronounced for galaxies
moving near edge-on. 

\begin{figure}
 \centering\resizebox{\hsize}{!}{\includegraphics[angle=0]{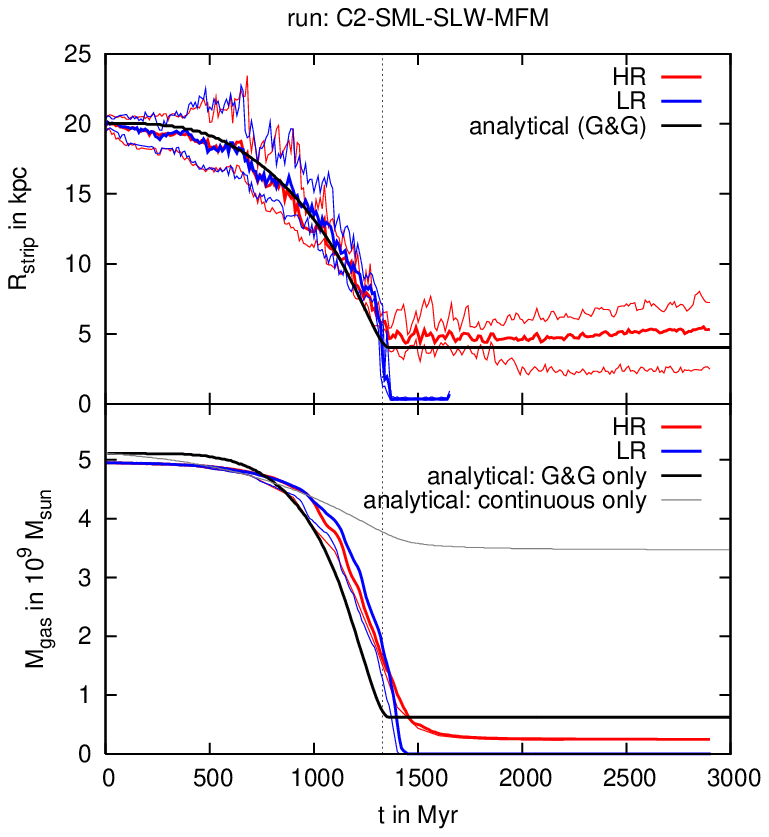}}
 \caption{Same as Fig.~\ref{fig:comp_an_num}, but comparing the high
   resolution (HR) and low resolution (LR) run for case C2-SM-SLW-MFM.}
 \label{fig:comp_an_num_res1}
 \end{figure}

For the evolution of the galaxy's gas disc mass and radius, two consequences
arise from this effect: 

Firstly, in cases where the galaxy moves with medium
to face-on inclination, not only the central density, but also the central
projected surface density, $\Sigma$, decreases, though much milder. The decrease in
surface density makes the gas disc more vulnerable to ram pressure pushing (compare to
Eq.~\ref{eq:fgrav}). For quite strong ram pressures as in run C2-SM-SLW-MFM,
this could cause the galaxy centre to be stripped prematurely. This indeed
happened in the low resolution case of run C2-SM-SLW-MFM. 
 In the high resolution case, the gas loss behaves as
expected. Fig.~\ref{fig:comp_an_num_res1} shows the evolution of mass and radius
for the two different resolutions. 
%
%
Up to pericentre passage, the two runs are nearly identical. The only other
``critical'' cases in the sense described above are the runs C1-SM-FST-MF and
C3-SM-FST-MF. For both runs, the gas disc radius and mass evolution agrees
well with the analytical estimate up to shortly before pericentre
passage. In both runs, the galaxy is stripped completely shortly before
pericentre passage, which is slightly faster than predicted by the analytical
estimate. We conclude that this sudden gas loss shortly before pericentre
passage is due to the numerical effect described above. However, we also
conclude that the galaxies
would indeed be stripped completely as the one in run C1-SM-FST-E is stripped
completely but is not expected to suffer strongly from the numerical
viscosity, as shown in Fig.~\ref{fig:comp_prof}.

\begin{figure}
 \centering\resizebox{\hsize}{!}{\includegraphics[angle=0]{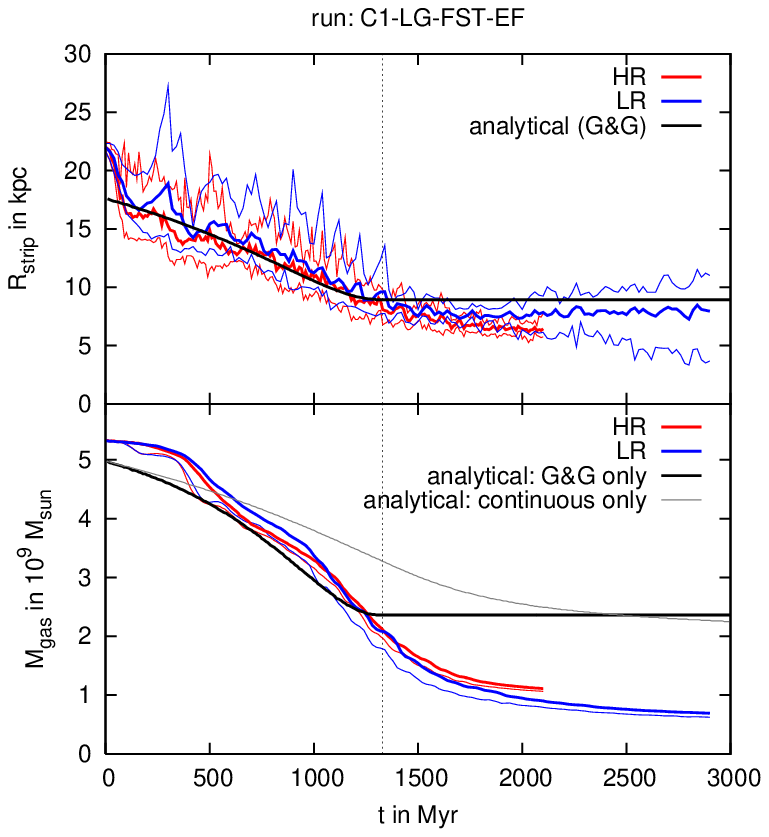}}
 \caption{Same as Fig.~\ref{fig:comp_an_num_res1}, but comparing the high
   resolution (HR) and low resolution (LR) run for case C1-LG-FST-EF.}
 \label{fig:comp_an_num_res2}
 \end{figure}

\begin{figure*}
 \centering
\resizebox{0.49\hsize}{!}{\includegraphics[width=\textwidth]{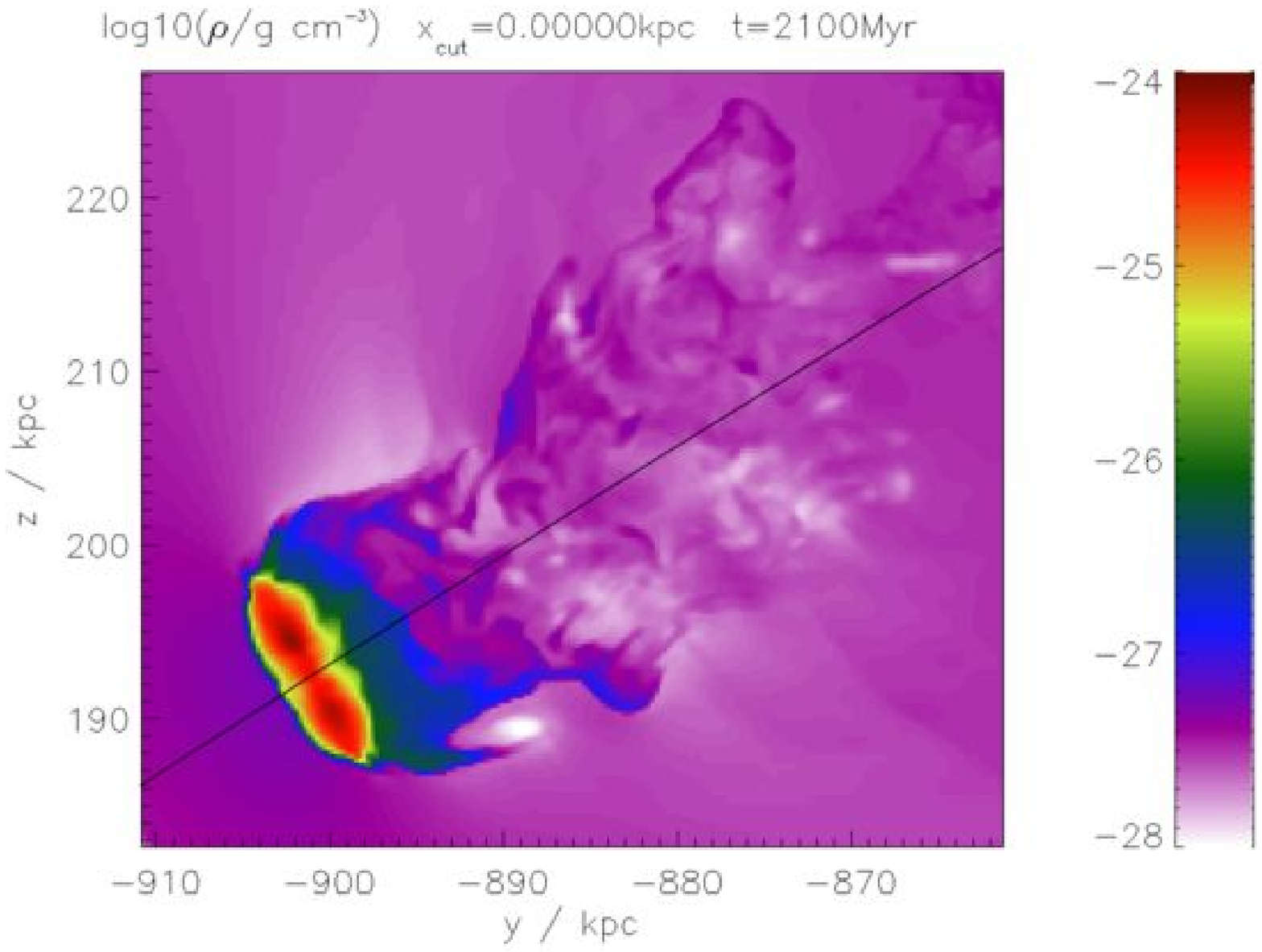}}
\resizebox{0.49\hsize}{!}{\includegraphics[width=\textwidth]{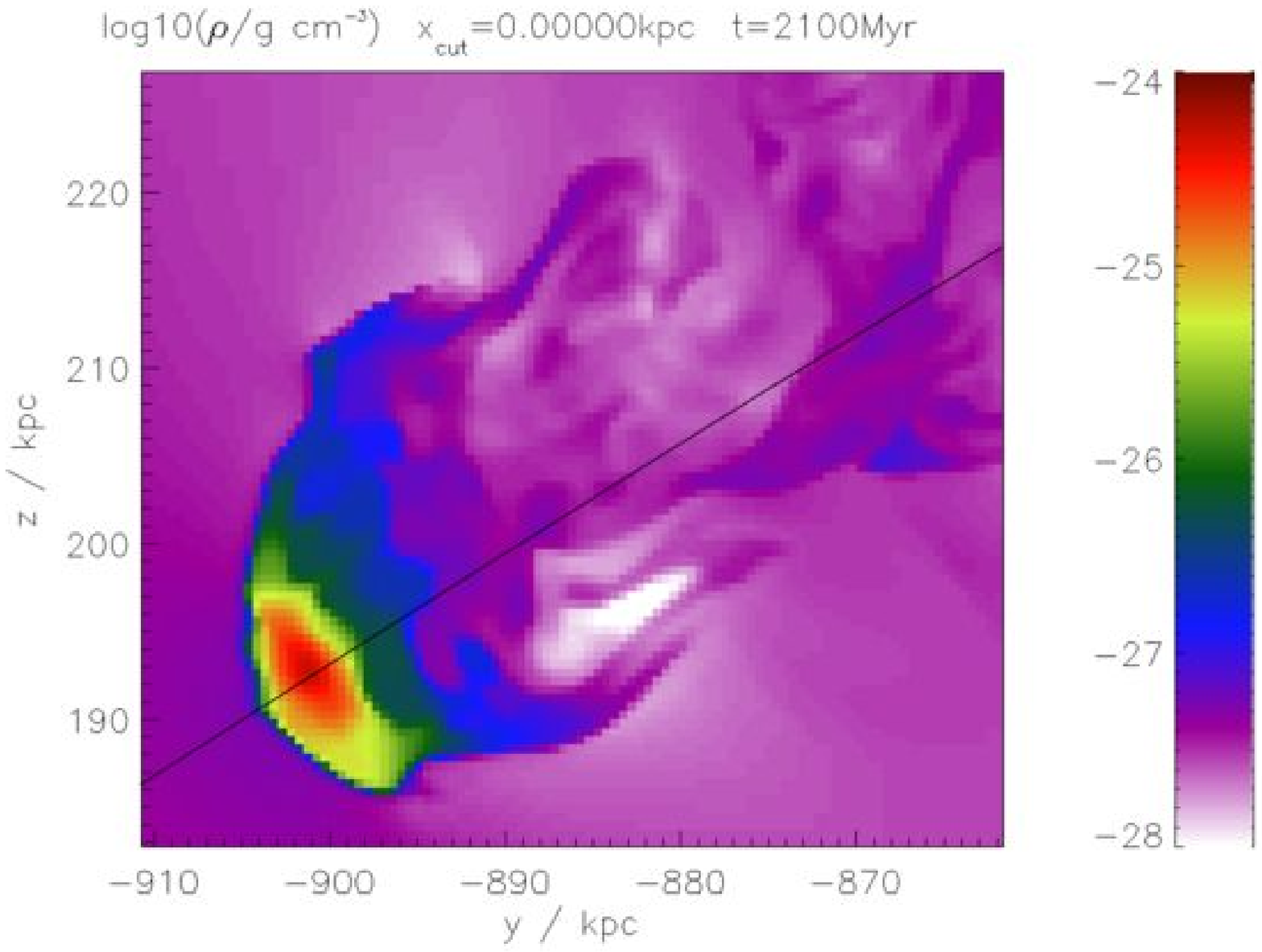}}
 \caption{Colour-coded gas density in the
   orbital plane for run C1-LG-FST-EF, at time $t=2100\Myr$. The left panel is for high resolution
   (HR), the right panel for low resolution (LR). The black line marks the
   galaxy's orbit.}
 \label{fig:slice_dens_res}
 \end{figure*}

The second consequence concerns continuous
stripping. Figure~\ref{fig:comp_an_num_res2} compares the evolution of the
mass and radius for run  C1-LG-FST-EF for two different resolutions. 
%
%
Up to $\sim 1500\Myr$, both results agree very well. After that, the low
resolution run yields a higher mass loss rate than the high resolution
run. This is due to the fact that the influence of the numerical diffusion
on the shape of the remaining gas disc is no longer
negligible, as can be seen in Fig.~\ref{fig:slice_dens_res}.
%

Concerning the evolution of the mass and radius of the remaining gas disc, we
conclude that the results presented in Fig.~\ref{fig:comp_an_num} are reliable
at least up to pericentre passage. After pericentre passage, the mass loss
rates due to continuous stripping derived by our simulations, are upper limits.


%
\bibliographystyle{mn2e}
\bibliography{%
../../BIBLIOGRAPHY/theory_simulations,%
../../BIBLIOGRAPHY/hydro_processes,%
../../BIBLIOGRAPHY/numerics,%
../../BIBLIOGRAPHY/observations_general,%
../../BIBLIOGRAPHY/observations_clusters,%
../../BIBLIOGRAPHY/observations_galaxies,%
../../BIBLIOGRAPHY/galaxy_model,%
../../BIBLIOGRAPHY/gas_halo,%
../../BIBLIOGRAPHY/icm_conditions,%
../../BIBLIOGRAPHY/clusters,%
../../BIBLIOGRAPHY/else}

\bsp

\label{lastpage}

\end{document}